# Uncovering Material Deformations via Machine Learning Combined with Four-Dimensional Scanning Transmission Electron Microscopy

Chuqiao Shi[1], Michael C. Cao[1,2], Sarah M. Rehn[3], Sang-Hoon Bae[4,5], Jeehwan Kim[6], Matthew R. Jones[1,3], David A. Muller[2,7], Yimo Han[1*]

[1] Department of Materials Science and NanoEngineering, Rice University, Houston, TX

[2] School of Applied and Engineering Physics, Cornell University, Ithaca, NY

[3] Department of Chemistry, Rice University, Houston, TX

[4] Department of Mechanical Engineering and Materials Science, Washington University in Saint Louis, Saint Louis, MO

[5] Institute of Materials Science and Engineering, Washington University in Saint Louis, Saint Louis, MO

[6] Department of Mechanical Engineering, Massachusetts Institute of Technology, Cambridge, MA

[7] Kavli Institute for Nanoscale Science, Cornell University, Ithaca, NY

[*] To whom correspondence should be addressed:
Yimo Han (Email: yimo.han@rice.edu)

ABSTRACT

Understanding lattice deformations is crucial in determining the properties of nanomaterials, which can become more prominent in future applications ranging from energy harvesting to electronic devices. However, it remains challenging to reveal unexpected deformations that crucially affect material properties across a large sample area. Here, we demonstrate a rapid and semi-automated unsupervised machine learning approach to uncover lattice deformations in materials. Our method utilizes divisive hierarchical clustering to automatically unveil multi-scale deformations in the entire sample flake from the diffraction data using four-dimensional scanning transmission electron microscopy (4D-STEM). Our approach overcomes the current barriers of large 4D data analysis and enables extraction of essential features even without a priori knowledge of the sample. Using this purely data-driven analysis, we have uncovered different types of material deformations, such as strain, lattice distortion, bending contour, etc., which can significantly impact the band structure and subsequent performance of nanomaterials-based devices. We envision that this data-driven procedure will provide insight into materials' intrinsic structures and accelerate the discovery of novel materials.

1. INTRODUCTION

Recent advances in the synthesis of materials have led to novel material structures at the nanometer scale. Atomic structure and deformations in these nanomaterials determine their chemical, electronic, and optical properties, affecting their efficiency and performance in their targeted applications. For example, novel epitaxial growth of optically tunable heterostructures in III-V semiconductors[1–3] and two-dimensional (2D) materials[4–8] can lead to local strain and dislocations, which greatly affect the electronic and optical properties due to changes in the local band structure. Even within nanomaterials comprised of a single crystal, such as structurally designed anisotropic metallic nano-plates or prisms[9–11], the local bending or deformations also play an important role in determining their optical responses and catalytic behaviors. To study these fine features in the materials, conventional high-resolution transmission electron microscopy (HRTEM) and annular dark field scanning transmission electron microscopy (ADF-STEM) have been utilized to reveal the local atomic structure [2,3,7]. While the limits of the electron microscope resolution are constantly being pushed[12], a common restriction on these imaging techniques is the limited field of view in the sample. As studying the overall structural information of the materials is crucial for mass production and large-scale processing for applications in next-generation devices, techniques such as nanobeam electron diffraction[13–15] that can map large sample areas have attracted tremendous attention for their potential in determining the sample structure on a large scale.

Although nanobeam electron diffraction has been used for decades to acquire electron diffraction patterns from a large sample area, this technique was limited by conventional charge-coupled device (CCD) detectors, which are too slow to collect detailed structural information across the entire sample. However, the development of fast direct electron detectors[16] now allows the collection of a momentum-resolved nanobeam diffraction pattern at each scanning position in a STEM experiment (Fig. 1a), thus generating four-dimensional (4D) data (Fig. 1b). Therefore, the technique enabled by these detectors is often referred to as 4D-STEM[17]. Among the direct electron detectors, the electron microscope pixel array detector (EMPAD)[18] has high dynamic range and sensitivity, capable of recording quantitative diffraction patterns without saturating the center beam or cutting off the weak diffracted spots. Using the EMPAD in scanning nanobeam diffraction mode, strain profiles in 2D materials have been mapped across a micrometer scale with sub-picometer precision[19]. However, this approach relies heavily on prior knowledge of the sample structure, as well as the accuracy of identifying each diffraction disk and determining their centers in the diffraction patterns. The large quantity of data from diverse samples also makes it difficult to utilize 4D-STEM for newly synthesized materials with unexpected lattice deformations, which have a large impact on material properties and device performance.

Recently, machine learning has emerged as a promising method applied in microscopy [20–25] due to its capability in analyzing complex patterns in large datasets. Specifically, unsupervised learning, which does not require training data, has been utilized to identify the stacking order[26] and twin boundaries[27] in materials. To further extend unsupervised learning for deformation and fine structure study, we utilized a divisive hierarchical unsupervised clustering architecture for rapid and semi-automatic 4D data analysis and feature mapping according to intrinsic characteristics and similarity (Fig. 1c). This approach has greatly improved the speed and accuracy of uncovering unexpected but significant fine structures and deformations in the materials.

## 2. METHODS

The process's overall scheme is described in Fig. 1. After the data acquisition, three main steps are used to process the 4D data: preprocessing the diffraction patterns, hierarchical clustering of the data, and visualizing the results.

**2.1 Preprocessing**

Preprocessing involves aligning and masking the diffraction patterns. Alignment corrects the drift in the diffraction patterns caused by slightly misaligned beam tilt when scanning large areas (Fig. S1). To do that, we align the center of mass (CoM) of the center beam. We added a circle mask in the center of each diffraction pattern to select the center beam (Fig. S1b). Then, the CoM of the center beam was calculated with sub-pixel precision, followed by moving the diffraction pattern towards the center of the detector (Fig. S1a). We repeated these three steps until the standard deviation of the CoM in all scanning positions, as well as the error between the CoM and the center of the detector, becomes smaller than 0.01 pixel. This step avoids the confusion caused by the translational shift of the diffraction patterns due to reasons other than the intrinsic sample structure.

Masking contains two parts, where the first part uses a ring mask to mask out the low-angle scattering as well as the zero-padded regions caused by the alignment. We determined the inner and outer radius through plotting the standard deviation (STD) of diffraction pattern. In the example data (Fig. S2a,e), the bright center beams always show highest STD due to the high intensity. From the rotational STD plots (Fig. S2b,f), we were able to identify the diffraction area and set the inner/outer radius to block the background and the noise (Fig. S2c,g). The second part selects the diffraction regions where the crystal information are stored. In this part, we select the features based on the STD of the intensities in the remaining area. The regions with high STD (30% of the highest) among the ring-masked dataset are selected in each diffraction pattern (Fig. S2d, S2h) and flattened as a feature vector, which is used in the following cluster analysis. The preprocessing helps organize the 4D datasets so that our method will mainly extract structural information from the datasets, while avoiding any ambiguity caused by the microscope misalignment and background.

**2.2 Hierarchical clustering**

The hierarchical clustering architecture is illustrated in Fig. 1c. A single round of clustering is insufficient to determine all the features in the 4D dataset due to features at varying length scales. Initial clustering separates large scale features, typically different materials in the sample, since they cause more noticeable differences in the diffraction pattern. However, diffraction patterns within these initial clusters then have more subtle differences caused by small-scale features, which cannot be separated through a single clustering round. To extract these features at different scales, we employ a divisive hierarchical clustering architecture to cluster the diffraction patterns in the 4D dataset. The divisive architecture is a "top-down" approach, which starts from the whole dataset, and then clustering is performed recursively when moving down the hierarchy.

In a single round of clustering, we compared different unsupervised learning methods and identified that K-means[28] shows the optimal performance considering combined accuracy and speed (Table S1 and Fig. S3). The K-means algorithm is a common clustering method that divides our real-space points, each represented by a feature vector, into *K* clusters through minimizing the variance (squared Euclidean distance) within each cluster. To automatically determine the number of clusters, or the *K*, especially when little prior knowledge about the datasets was presented, we utilize the *elbow method* [29] to determine the *K* number in each round clustering (Fig. S4). In the *elbow method*, we calculated the total *within-cluster sum of squares (WSS)* and plotted the error curves according to the number of clusters (or *K*). Since we do not need precise *WSS* values, we adopted the faster but negligibly less precise Mini-Batch K-means method (Fig. S3e) to determine the *elbow point*.

## 2.3 Visualization

Finally, since each diffraction pattern represents one pixel in real space, the labels from the clustering results can be mapped back to real space for a better understanding of the material structures. As shown in the top panel of Fig. 1d, the clustering result is visualized in a real-space color-coded map, with each color representing a unique structural feature characterized in a cluster of similar diffraction patterns. Alternatively, we can also visualize the data distribution in a low-dimensional manifold structure (Fig. 1d bottom panel). The manifold structure is the projection of the high dimensional diffraction patterns to 3D space for visualization, which is achieved by the recently developed uniform manifold approximation and projection (UMAP)[30]. The lower dimensional manifold structures can assist in the visualization of the distribution and variance of the clusters, which can be used to determine the differences in major physical parameters. Through real-space and manifold structure visualization after clustering, we can map the fine deformations to better understand the distribution of structural features in the sample.

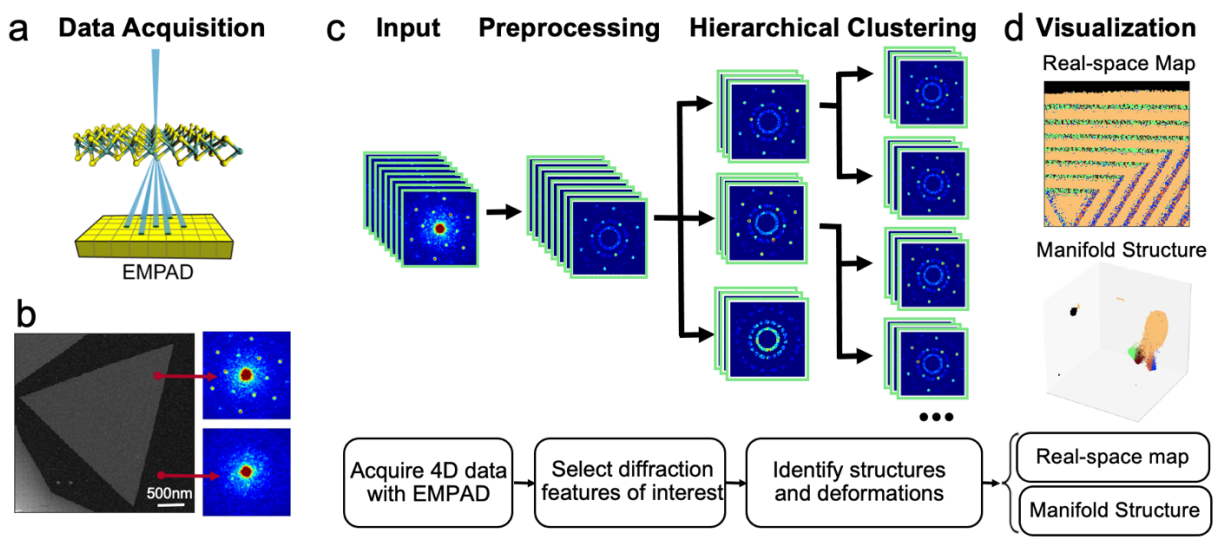

**Figure 1 | Schematic of unsupervised learning of 4D-STEM datasets.** (a) Schematic of the EMPAD operation, where a diffraction pattern is recorded at each scanning position. (b) A 4D dataset that contains a full diffraction pattern at each pixel in the real-space map. (c) Schematic of the divisive hierarchical clustering workflow on diffraction patterns in the 4D data. (d) Visualization of the clustering results, where the features are mapped back to real space (top) to determine their distribution and the diffraction patterns are projected onto a lower-dimensional manifold space (bottom) for visualization.

3. RESULTS
3.1 Deformations in $WS_2$-$WSe_2$ lateral heterojunction.

To test the accuracy our clustering architecture, we applied our approach to 4D datasets of 2D epitaxial lateral heterojunctions, which contain unique strain-engineered structures that enable tunable optical properties[19]. The sample is composed of an outer and inner region of $WS_2$ and a middle region of $WSe_2$, which is not apparent in its ADF-STEM image (Fig. 2a) as the tungsten atoms dominate the ADF contrast. During the clustering, structural features at different scales within the entire flake have been uncovered hierarchically. After three rounds, as shown in the dendrogram (Fig. 2b), fine deformations have been uncovered in the sample.

The real space map from the first two rounds of clustering is shown in Fig. 2c, where the method effectively separated the background, $WS_2$, and $WSe_2$ (Fig. S5a – S5c). The crystalline samples are distinguished from the amorphous substrate in the first round of clustering, which is intuitively obvious. In the second round, the $WSe_2$ and $WS_2$ are differentiated by the diffraction spots spacing caused by the lattice mismatch between $WS_2$ and $WSe_2$ (Fig. S5d). The real space map from unsupervised learning provides a precise interface between $WS_2$ and $WSe_2$ with defects (Fig. S6), which were not recognized in the ADF images. The two separated clusters of $WS_2$ and $WSe_2$ in the manifold structure (Fig. 2f) indicate that the lattice constant changes across the junction are discrete. We measured the lattice constant of each diffraction pattern in the junction sample, and the histogram (Fig. 2i) confirms the discrete change.

In the third round of clustering, the sub-clusters of $WS_2$ and $WSe_2$ from the second round are analyzed to reveal finer features. Our method separates the $WS_2$ cluster into two sub-clusters, where the real space map shows a rotational periodicity of the material with graded interfaces (Fig. 2d). The mean diffraction patterns (Fig. S5e and S5f) of each sub-cluster display a slight rotation in the reciprocal lattice (Fig. S5g), which corresponds to different lattice rotations in real space. Unlike the discretely separated $WS_2$ and $WSe_2$ clusters from the previous round, these two clusters are mixed in the manifold representation (Fig. 2g), indicating that the rotation angle changes continuously. We measured the rotation angle of each diffraction pattern in $WS_2$, and the histogram (Fig. 2j) confirms the capability of our method to identify continuous structural distortions in materials.

Meanwhile, the $WSe_2$ cluster is separated into four clusters, and the real space map is shown in Fig. 2e. The averaged diffraction patterns of each cluster differ in the intensity of the second order spots (Fig. S5h-k), caused by lattice tilts (or ripples) in the sample[31,32]. Specifically, three separate regions in the corners show different directional ripples since the second-order diffraction spots along one direction display a much stronger intensity than the other two directions (Fig. S5i-k). In contrast, the area close to the center is a flat region where all the second-order spots have similar intensity (Fig. S5h). The 3D manifold structure of the $WSe_2$ cluster (Fig. 2h) shows that the tilts in the ripple area are continuous. The results we achieved here are consistent with literature where the ripples form to relax the strain induced by $WS_2$-$WSe_2$ lattice mismatch[19].

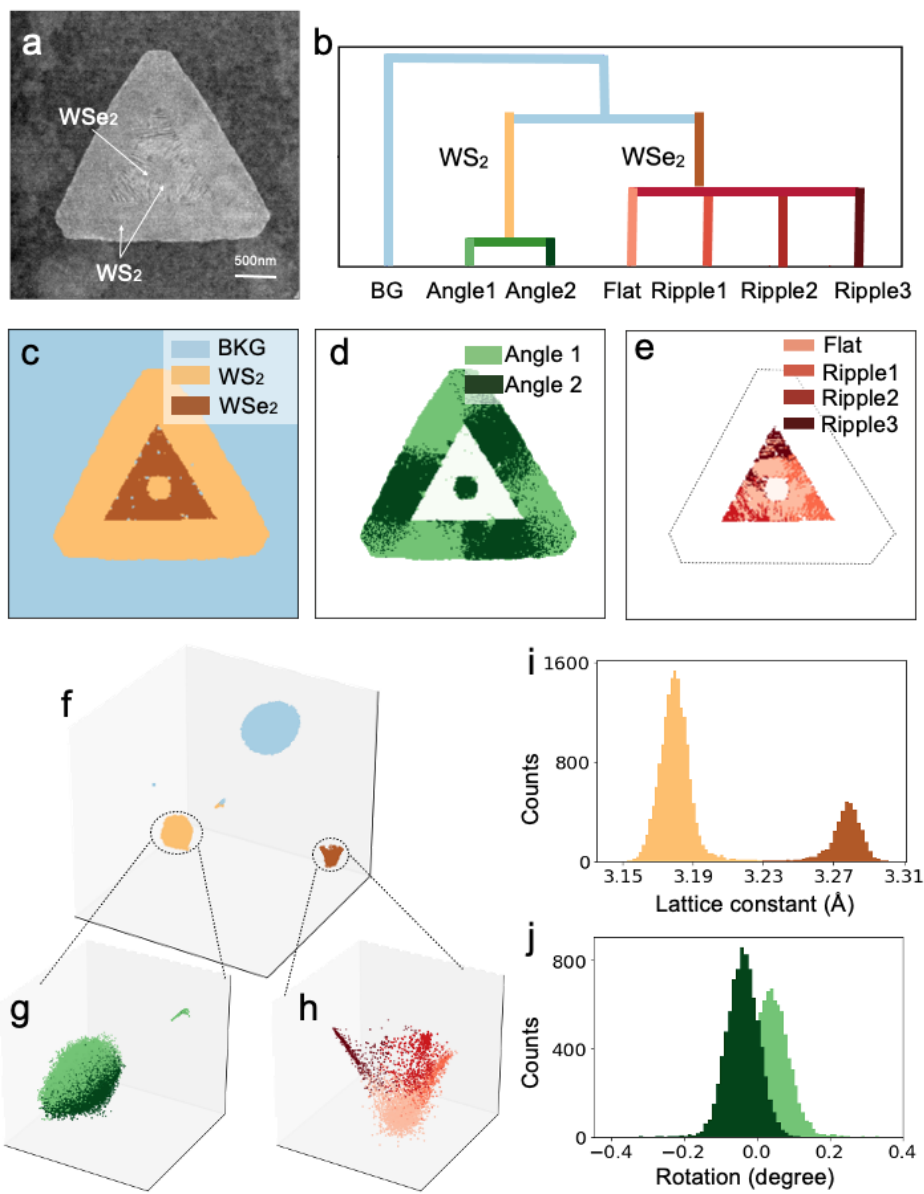

**Figure 2 | Clustering results on WS$_2$-WSe$_2$ lateral heterojunction.** (a) ADF-STEM image of the WS$_2$-WSe$_2$ junction. (b) Dendrogram of the unsupervised learning workflow. (c-e) Real space maps from the unsupervised learning results showing different compositions and deformations in the sample. (f-h) 3D manifold structure displaying the data distribution in each round of the clustering. (i) Histogram of measured lattice constants of WS$_2$ and WSe$_2$ color-coded with results in the second-round clustering. (j) Histogram of measured rotation angles of WS$_2$ color-coded with results in the third-round clustering of WS$_2$.

**3.2 Uncovering minor ripples in WS$_2$-WSe$_2$ superlattices**

To test how sensitive our method is for minor deformations in materials, we utilized a coherent 2D superlattice sample which contains a much smaller ripple structure with an aspect ratio of ~1/30 in narrow $WSe_2$ stripes[8]. Using our method on the entire superlattice sample (Fig. 3a), structural features at different scales were uncovered hierarchically. Our method identified different flakes in the first round of clustering (Fig. 3b). Then lattice differences between $WS_2$ and $WSe_2$ were illustrated in the second round (Fig. 3c). The next few rounds unveiled directional uniaxial strain (Fig. 3d and Fig. S7a-e), as well as minor ripples hidden in the $WSe_2$ (Fig. 3e and Fig. S7f-j). Due to the coherency in the superlattice, the ripples presented here are much smaller than what we observed in the $WS_2$-$WSe_2$ lateral heterojunction sample. The manifold structures of the zoomed-in dataset from each round of clustering provide a view of each cluster and subcluster in data space (Fig. S8).

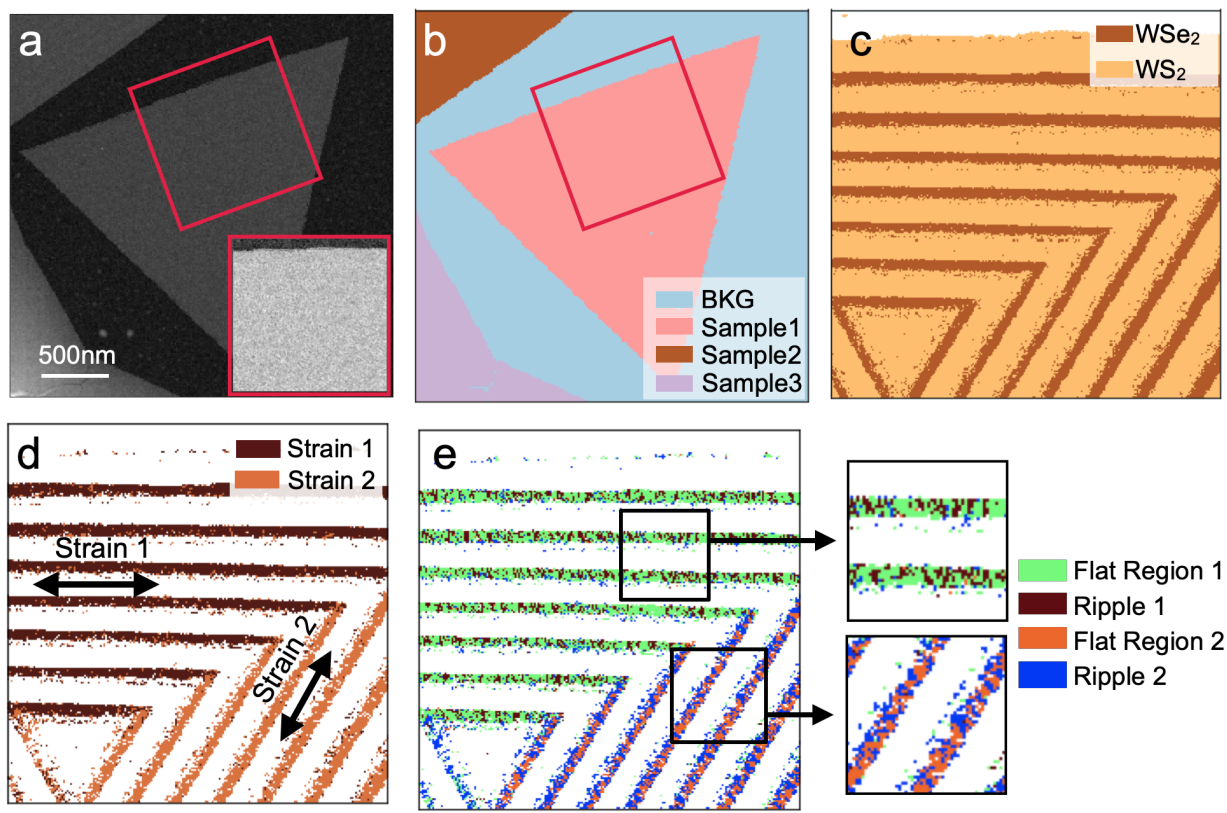

**Figure 3 | Clustering results on $WS_2$-$WSe_2$ superlattices.** (a) ADF-STEM images of the 2D multi-junction superlattice with a zoomed-in area shown in the inset. The ADF contrast is dominated by heavy tungsten atoms, thus no superlattice structure appears. (b) Real-space map of the clustering results from the 4D dataset of the $WS_2$-$WSe_2$ multi-junction sample, where different flakes are identified from the substrate. (c) Zoomed-in real-space maps of the multi-junction

sample area shown in the red box in (a). (d) Real-space map of the sub-clustering results in WSe$_2$ with two colors indicating different strain profiles. (e) Real-space map of another clustering round on each subcluster in (d), providing more structural details of lattice ripples in WSe$_2$ due to the strain.

**3.3 Uncovering bending contours in silver nanoprisms**

Unlike 2D materials, thicker samples usually present more complex deformations and lattice distortions, which result in more challenges in understanding their local structure. We use our method to investigate the deformations in silver nanoprisms, which have been widely studied due to their unique optical properties that show great potential in many applications[9–11]. The silver nanoprisms were drop casted on an amorphous carbon supporting film followed by air drying, which introduced internal deformations due to the surface tension. The ADF-STEM (Fig. 4a) shows an entire flake of a micrometer-sized silver nanoprism, which is known to be single crystal with the [111] zone axis perpendicular to the flat surfaces. Here, we show that our machine learning approach can elucidate deformations in nanoprism flakes in a facile and semi-automated way (Fig. S9).

Focusing on a corner of the nanoprism (Fig. 4b), the first round of clustering distinguishes the sample from the amorphous supporting film (Fig. S10a) due to the obvious difference between the amorphous film (Fig. S10b) and the nanoprism (Fig. S10c) in the diffraction patterns. The following round of clustering identify the contours in the nanoprism flake (Fig. S10d-f). To further investigate the deformation, we clustered the contour data into smaller sub-clusters and identified the two sides of the contour (yellow and orange in Fig. 4c), which is consistent with the general structure of a bending contour. From the averaged diffraction patterns (Fig. S10h and i), the difference between the two sides of the bending contour in diffraction space is the intensity variation in the conjugate diffraction spots, which can be explained as the incident electron beam approaching the atomic planes at different angles at two different bending sides. On each side, the Bragg condition in one direction is fulfilled, and the diffraction of the incident beam would result in strong corresponding diffraction peaks (Fig. S11a). In addition, by placing masks on the conjugate diffraction spots to form virtual dark-field (DF) images (Fig. S11b), the high contrast contours in the virtual DF images are consistent with the previously reported experimental DF-TEM images in silver nanoprisms[9].

In the manifold structure (Fig. 4d-f), the data is first segmented into two main clusters (Fig. 4d), amorphous background (black) and nanoprism (dark purple). Further sub-clustering on the nanoprism cluster separated the ring-like appendage (bending contour) from the original sample manifold as the new-subcluster (Fig. 4e). The thin asymmetric shape of the bending contour manifold (Fig. 4f) indicates a continuous deformation, which is shown in the diffraction pattern as an intensity change in the two conjugate diffraction spots.

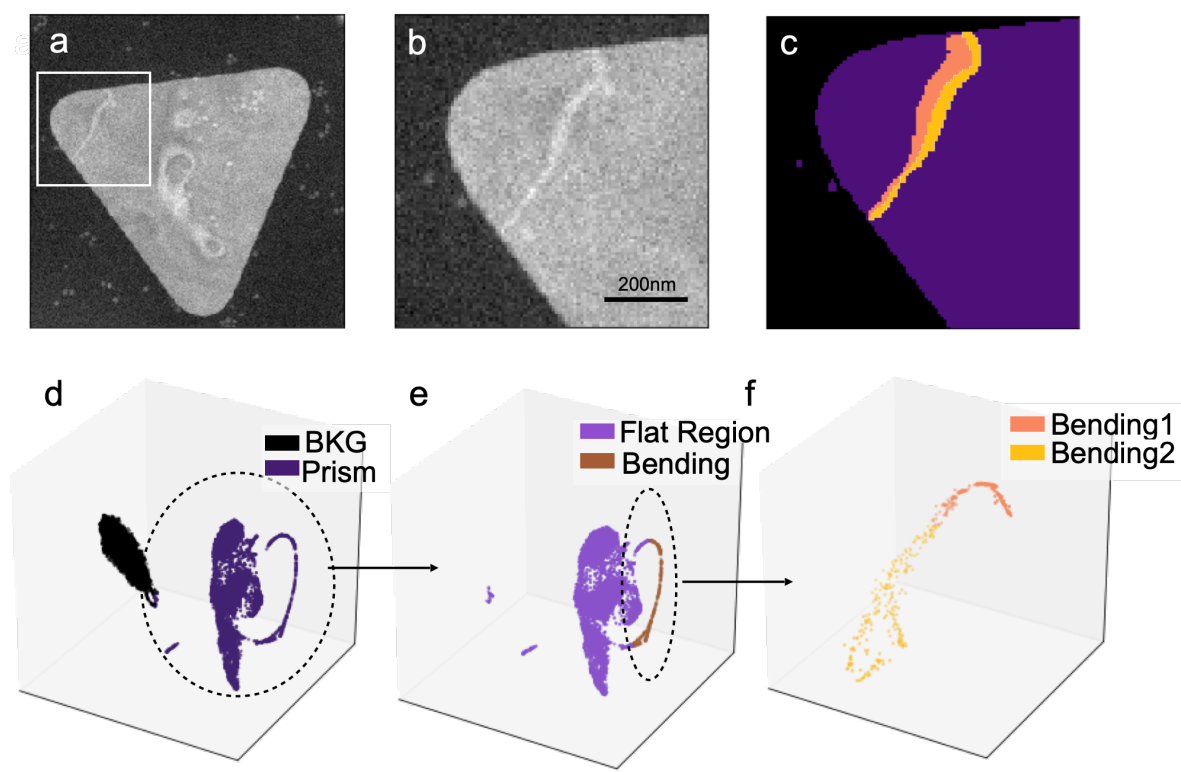

**Figure 4 | Clustering results of silver nanoprisms.** (a) ADF-STEM image of a silver nanoprism placed on amorphous $SiN_x$ supporting film. (b) Zoomed-in ADF image from the white box in (a). (c) Real space map of the final cluster and sub-cluster results. (d-f) Manifold structure of the data in each hierarchical clustering process.

**3.4 Clustering real space images for virtual imaging.**

Conventionally, we represent 4D-STEM datasets in real-space major order $(x, y, k_x, k_y)$. However, we can also view the data in momentum-space major order $(k_x, k_y, x, y)$. Instead of a 2D array of diffraction patterns, we can visualize the 4D dataset as a 2D array of real-space images generated from diffraction pattern intensities at a single momentum-space coordinate (Fig. 5a). Due to this property of 4D datasets, our hierarchical method could also be extended to clustering these real-space images (Fig. 5b), aimed at analyzing the data from a different dimension. However, the quantitative intensity of real-space images varies dramatically across the diffraction space. Consequently, the clustering results are dominated by the intensity effect, overpowering the actual feature information shown in the contrast (Fig. S12a). To uncover the actual features, we normalized the real space image as an additional preprocessing step before clustering, which provides similar intensity in the real space images for better recognition of sample features using our method. Followed by the hierarchical method, the real-space features in different momentum-space pixels can be uncovered.

We re-investigate 4D data of the $WS_2$-$WSe_2$ superlattices using this new clustering approach to segment the diffraction space into multiple clusters (Fig. 5c). This method works by effectively placing virtual objective apertures in the diffraction space, which can accurately select the diffraction coordinates based on their generated real-space image similarity. In each cluster, the averaged real-space image shows results equivalent to different modes in STEM, including virtual bright-field (BF) (Fig. 5d) and DF (Fig. 5e and 5f) images. The DF images of different flakes are separated (Fig 5e and 5f) according to their different lattice orientations (cyan and yellow in Fig. 5c). Compared with the hierarchical clustering on diffraction patterns (Fig. 3), the results from clustering real-space images provide crystal information that is stored in momentum-space. However, we observed striping in the BF image (Fig. 5d) generated from pixels in the center beam. Upon further investigation of the data (Fig. S12c,d), we found the stripes are caused by the center beam alignment in the pre-processing of the 4D data, which indicates the importance of more accurate alignment of the beam tilt for 4D nanobeam electron diffraction mode in STEM.

The virtual imaging through unsupervised learning can not only be used on 2D materials but also bulk materials. We investigated the epitaxial InGaP grown on GaAs, where the complexity of strain and dislocations in the InGaP layer create challenges to quantitatively understand the material structure from conventional methods (ADF-STEM image in inset of Fig. 5h). As the strain profile in the material dramatically affects its properties and performance, uncovering the complicated crystalline structure in the film becomes crucial. For such complex materials, our real-space clustering approach segments the reciprocal space into six parts. A BF image (Fig. 5i) is generated from the segment where the center beam is located (green part in Fig. 5h). A low-angle ADF (LAADF) image (Fig. 5j) is reconstructed from low scattering angles between the center beam and first order diffraction spots (dark blue part in Fig. 5h), which highlights the amorphous carbon protecting layer on top of the film due to the amorphous scattering ring formed at this low angle. In addition, higher angle areas that exclude the diffraction spots (light blue part of Fig. 5h) display elastically scattered electrons that roughly represent the thickness of the sample (Fig. 5k). The remaining three clusters indicate the crystallinity of the thin film, providing virtual DF images of three lattice orientations (Fig. 5l-5n). From the diffraction map (Fig. 5h), we conclude that Fig. 5l-m show small tilts possibly caused by the lattice strain in InGaP, while Fig. 5n displays a twin domain formed in the film.

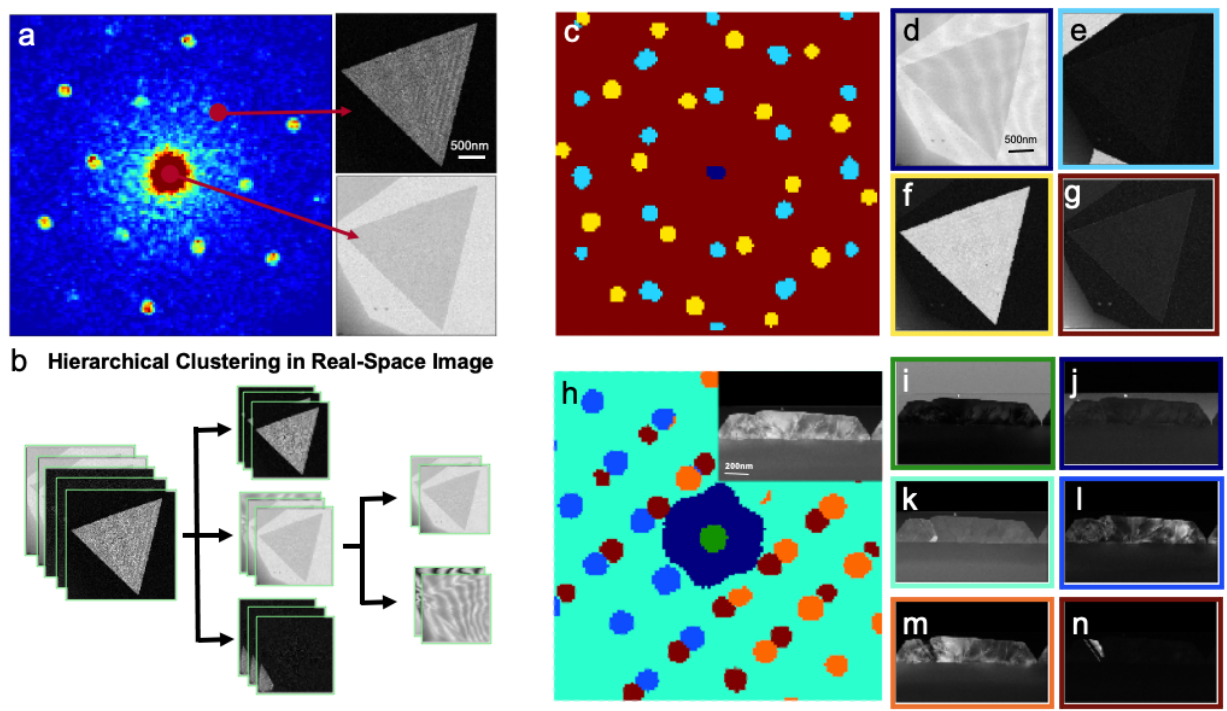

**Figure 5 | Clustering results on real space images in 4D-STEM datasets.** (a) Visualizing 4D data in a momentum-major order, where each pixel in diffraction pattern can be considered as a real-space image. (b) Schematic of the divisive hierarchical clustering architecture on real space images. (c) Map of hierarchical clustering results in diffraction space on a $WS_2$-$WSe_2$ superlattice. (d-g) Mean real-space images of the superlattice in each cluster. (d) displays the image from dark

blue area in (c), which represents a virtual BF image. (e) and (f) are the images from light blue and yellow portions in (c), corresponding to DF images for different flakes. (g) sums all other areas in (c), indicating a thickness variation in the sample. (h) Map of hierarchical clustering results in diffraction space on a cross-sectional InGaP/GaAs crystal, with an ADF image displayed in the inset. (i-n) Mean real-space images of the sample in each cluster. (i) corresponds to the center beam, the green area in (h), and shows a virtual BF image. (j) and (k) are from the dark blue and cyan area in respect, which are from the amorphous carbon and thickness effect in the sample. (l-n) are virtual DF images for the cross-section, showing strain effects and twin grains.

## 4. DISCUSSION

### 4.1 Dimension Reduction Methods Comparison

The EMPAD records 128x128 diffraction patterns so each pattern has 16834 pixels. However, the useful structural information is concentrated only in the diffraction spots, which is a small part in the whole pattern. A condensed representation of the data significantly improves computational speed and feature selection. In the preprocessing step, we chose to place STD masks on diffraction patterns to reduce the data size. However, there are more elegant approaches to reduce the dimension of the data such as Principal Component Analysis (PCA) and UMAP. But they are not desired for our hierarchical clustering method.

PCA computes the principal components based on matrix factorization but since it is applied to the whole dataset, only the main difference is captured while the minor differences are ignored. For example, when applying the hierarchical clustering on the PCA components of the $WSe_2$-$WS_2$ junction in Fig. 2, the $WSe_2$ and $WS_2$ samples can be separated in the first two rounds (Fig. S13b). However, in the next round of clustering, the rotations cannot be captured (Fig. S13c) unless we re-apply PCA on each sub-cluster before each round clustering, which takes more time and memory. The need to re-apply dimension reduction on each sub-cluster makes PCA undesirable for our clustering method.

The other feature extraction method we tested was UMAP, which approximates the features in low-dimensional manifold space. Compared with PCA, UMAP can capture both the major and minor features in the sample. However, the clustering results show striping artifacts (Fig. S13d), while the real deformation features are hidden. Due to this artifact, we don't use UMAP as a dimension reduction method but only as a visualization method.

**4.2 Clustering Methods Comparison**

There are alternative clustering methods besides K-Means, including Agglomerative clustering, BIRCH clustering, Spectral clustering, etc. To quantitively compare the efficiency and performance of different clustering methods, we created the ground truth label on the $WS_2$ cluster from the manually measured rotation map (Fig S3a). We set the positive rotation angle as label 1 and the negative rotation angle as label 2. Then we cluster this dataset with different methods, record the time and calculate the accuracy. Based on the results shown in Fig S3, the Mini-batch K-means method is the fastest. However, due to the random initialization of this algorithm, the clustering results are not stable. The spectral clustering provides the best accuracy, but takes a much longer time. Considering the tradeoff between time and performance, K-means was chosen as the clustering method in each round of the hierarchical clustering architecture.

Here, we summarize the combination of different dimension reduction and clustering methods on the $WS_2$ dataset. The time and accuracy of each method are shown in Table S1, which proved the STD selection and K-Means are suitable for our hierarchical clustering architecture.

**5. CONCLUSION**

We have demonstrated a method using divisive hierarchical unsupervised machine learning to accelerate and automate the study of lattice deformations in novel materials. As understanding lattice deformations will play an important role in the material properties and device performance, our method represents a crucial step towards a deeper and easier method for the analysis of subtle atomic features. We have applied this method to extract such features from different 2D lateral heterojunctions, thicker 3D materials, and cross-sectional crystals. The purely data-driven analysis uncovers different types of material deformations in the samples, such as strain, lattice distortion, bending contour, etc. This method may be potentially expanded to broader material systems or other imaging techniques that generate large and multidimensional datasets, benefiting the development of new materials, techniques, and applications.

**6. ACKNOWLEDGEMENTS**

C.S. and Y.H. are supported by start-up funds provided by Rice University. Y.H. acknowledge the support from the Welch Foundation (C-2065-20210327). M.C. and D.A.M is supported by the NSF MRSEC program (DMR-1719875). S.M.R. would like to acknowledge financial support from a National Science Foundation Graduate Research Fellowship (No. 1842494). This work made use of the Electron Microscopy Center at Rice University. We thank Mengnan Zhao and Guanhui Gao for useful discussion. We also thank Jiwoong Park and Saien Xie for providing 2D junction samples, which have been previously reported in these references [8,19].

## 7. AUTHOR CONTRIBUTIONS

Experiments and data analysis were performed by C.S. under the supervision of Y.H.; C.S. and M.C. contributed to the machine learning algorithm under supervision of Y.H.; samples were provided by S.M.R., M.R.J., S-H.B., J.K.; datasets in Fig. 1 to 3 were acquired at Cornell by Y.H. under the supervision of D.A.M.

## COMPETING INTERESTS

The authors declare that they have no competing interests.

## DATA AVAILABILITY

All data needed to evaluate the conclusions in the paper are present in the paper and/or the Supplementary Materials.

## CORRESPONDING AUTHORS

Correspondence to Yimo Han (Email: yh76@rice.edu)

**Supporting Information for**

**Uncovering Material Deformations via Machine Learning Combined with Four-Dimensional Scanning Transmission Electron Microscopy**

Chuqiao Shi[1], Michael C. Cao[1,2], Sarah M. Rehn[3], Sang-Hoon Bae[4,5], Jeehwan Kim[6], Matthew R. Jones[1,3], David A. Muller[2,7], Yimo Han[1*]

[1] Department of Materials Science and NanoEngineering, Rice University, Houston, TX

[2] School of Applied and Engineering Physics, Cornell University, Ithaca, NY

[3] Department of Chemistry, Rice University, Houston, TX

[4] Department of Mechanical Engineering and Materials Science, Washington University in Saint Louis, Saint Louis, MO

[5] Institute of Materials Science and Engineering, Washington University in Saint Louis, Saint Louis, MO

[6] Department of Mechanical Engineering, Massachusetts Institute of Technology, Cambridge, MA

[7] Kavli Institute for Nanoscale Science, Cornell University, Ithaca, NY

[*] To whom correspondence should be addressed:

Yimo Han (Email: yimo.han@rice.edu)

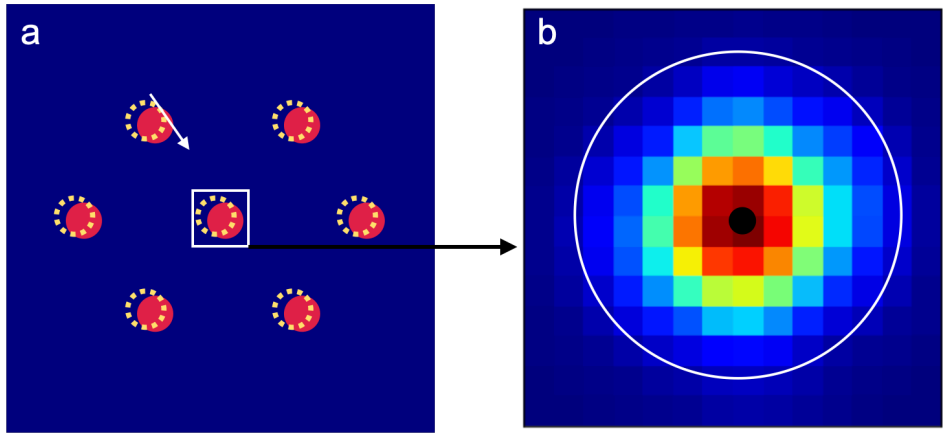

**Figure S1**. **Aligning diffraction patterns in preprocessing.** (a) Schematic of alignment in the preprocessing step. This step corrects the translational shift (indicated by the white arrow) of the diffraction patterns across the 4D mapping area caused by beam tilt in the microscope. Specifically, we correct this misalignment by aligning the center of mass (CoM) of the center beam (shown in the white box). We iterate this operation until the standard deviation for all CoM measures remains below 0.01 pixel in both $k_x$ and $k_y$ for all scanning positions in real space. (b) An example of the center beam in a diffraction pattern. We place a mask to select the center beam (white circle) and calculate the CoM (indicated by the black dot).

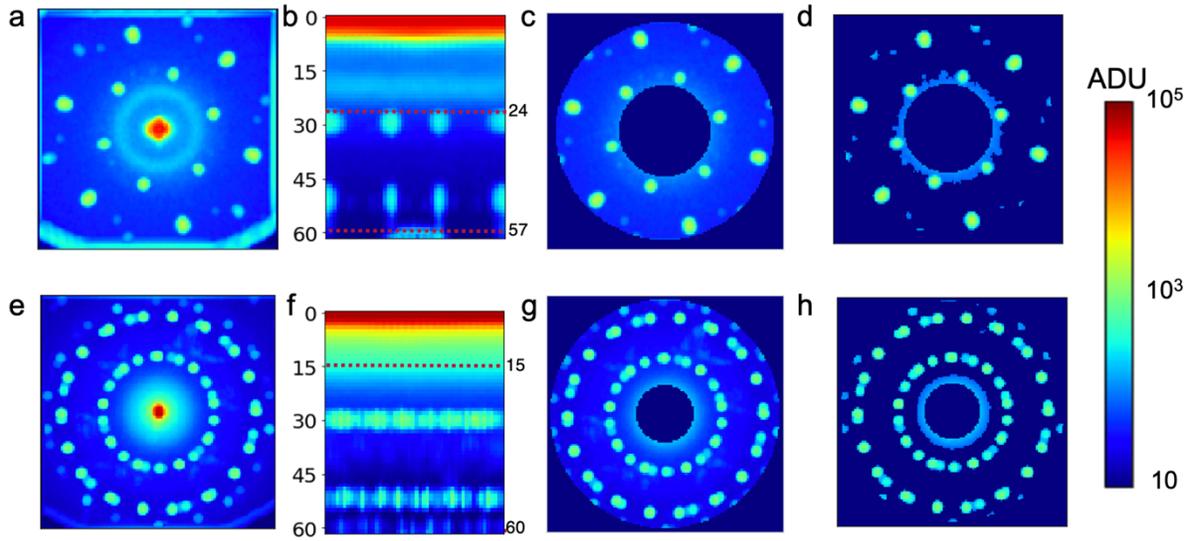

**Figure S2. Placing masks to select the diffraction areas of interest**. (a) Standard deviation diffraction image by plotting the standard deviation of the pixel intensity across all scanning points in real space. This diffraction image was from the $WS_2$-$WSe_2$ junction 4D data shown in Fig. 2. (b) Diffraction image from (a) plotted in (θ, r) coordinate, where the amorphous ring structure becomes a line for easy determination of the inner radius of the ring mask to block it. The inner and outer radii for the mask are indicated by the red dashed lines, which are 24 and 57 in pixels respectively. (c) Diffraction image in (a) with the mask placed on it. The same mask will be applied on all actual diffraction patterns in the 4D data to select the diffraction area of interest. (d). Diffraction image in (a) with the standard deviation mask placed on it. The remaining pixels (higher than 30% of the highest STD) will be flattened as a vector. (e) Standard deviation diffraction pattern, plotted in the same way as (a), of the 4D data collected on $WS_2$-$WSe_2$ superlattices described in Fig. 3. (f) Diffraction image from (e) in (θ, r) coordinate system with the inner and outer radii determined as 15 and 60 pixels. The amorphous ring in this dataset is much weaker due to the thin layer of the supporting film in this sample. (g) Diffraction image in (e) with the mask. (h) Diffraction image in (g) with the standard deviation mask placed on it.

Accuracy and Time of Clustering and Dimension Reduction Techniques

|  | **Std mask** | **PCA 3** | **UMAP 3** | **Raw** |
|---|---|---|---|---|
| **Agglomerative Clustering** | 77.4%, 31s | 51.4%, 7.8s | 77.6%, 7.6s | 77.5%, 140s |
| **BIRCH Clustering** | 77.6%, 35s | 52.5%, 1.5s | 75.6%, 1.6s | 77.3%, 127s |
| **Mini-Batch K-Means** | 83.2%, 0.05s | 53.3%, 0.03s | 80%, 0.02s | 83.3%, 1.89s |
| **Spectral Clustering** | 84.6%, 36s | 52.1%, 31s | 81%, 32s | 84.7%, 48s |
| **K-means** | **84.3%, 0.4s** | 53.5%, 0.09s | 80.4%, 0.11s | 83.9%, 32s |

**Table S1.** Comparison of dimension reduction methods & clustering methods. Standard deviation mask, PCA and UMAP are used to reduce the dimension and extract the features from the raw data before clustering, then five widely used clustering methods are applied on the extracted features. The PCA can only capture the dominant feature, but it cannot extract the minor features, such as the small rotation angles in WS2 sample. Its accuracy is the lowest and the labels are shown in Figure S14c. UMAP can capture the features, but the clustering labels (Fig. S14d) shows fake stripes caused by alignment in the preprocessing. The standard deviation mask method provides the same accuracy compared with the raw data across each clustering method, but dramatically reduces the time.

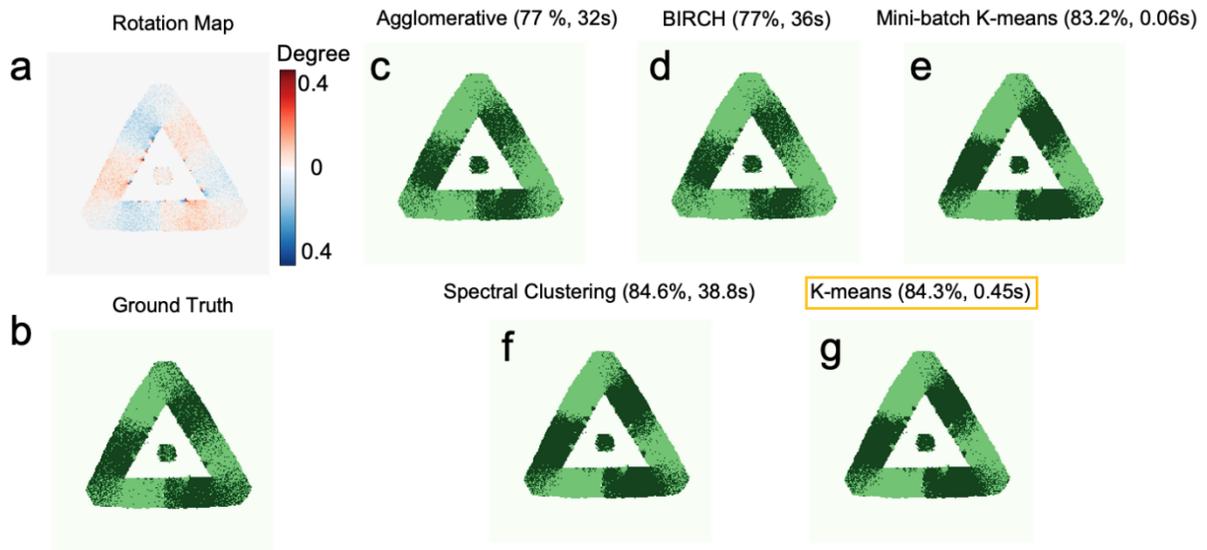

**Figure S3. Comparison among five widely used clustering methods.** (a) Rotation map measured manually, which has been reported in previous reference[1] (b) Ground truth labels of two rotation angles in $WS_2$ sample. According to the rotation map, positive angles are set as label 1 and negative angles are set as label 2. (c – g) Clustering results of five widely used clustering methods. Clustering time and accuracy are labeled on each panel. Mini-batch K-means method is the fastest, so we use it in the elbow method to determine the cluster number. However, since the batch selection is random in this algorithm, the clustering results are not stable. The spectral clustering provides the best accuracy but takes a much longer time. Considering the tradeoff between efficiency and performance, K-means is chosen as the clustering method in each round of the hierarchical clustering architecture.

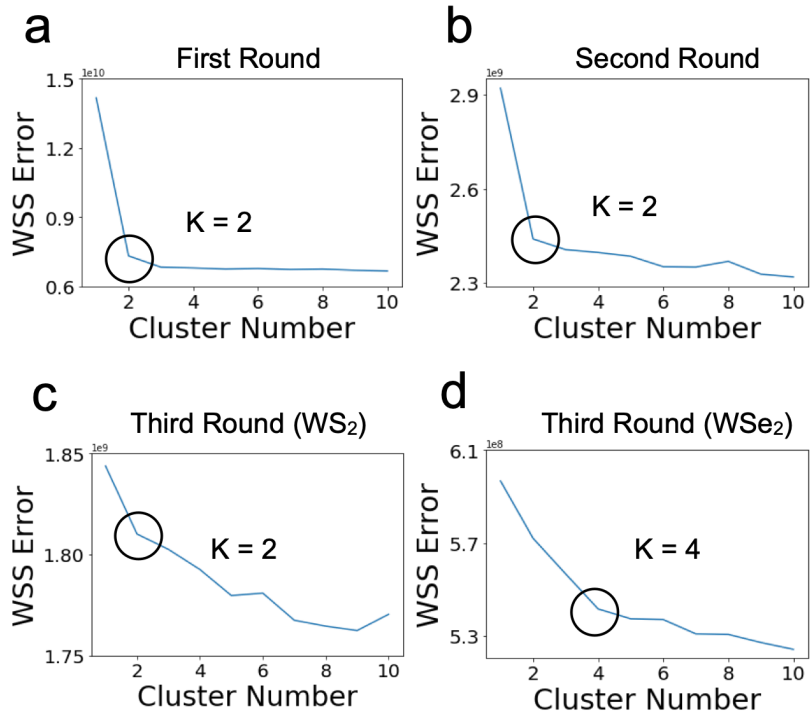

**Figure S4. Elbow method for determining the clustering number for each round.** Elbow method runs the clustering algorithms several times with different K numbers and calculates the Within-Cluster Sum of Squares (WSS) error each time. The WSS error trends downward with increasing cluster number. Afterwards, we determine the K number through the elbow point on the WSS curves. The elbow point is the inflection point where the second derivative of the WSS curve becomes negative, which indicates that adding one more cluster from K to K+1 will not reduce the error as much as adding one cluster from K-1 to K. The elbow method was highly efficient at selecting the optimum number of clusters for the various samples we analyzed and automated our methods for the 4D data processing. (a) The result from the first-round clustering of the junction sample in Fig. 2, where K=2 was selected as the elbow point. (b) The second-round clustering of the junction area, resulting in a clustering number of 2. (c) The third-round sub-clustering of $WS_2$ gives an optimal cluster number of 2. (d) The third-round sub-clustering of $WSe_2$ displays a K=4 elbow point.

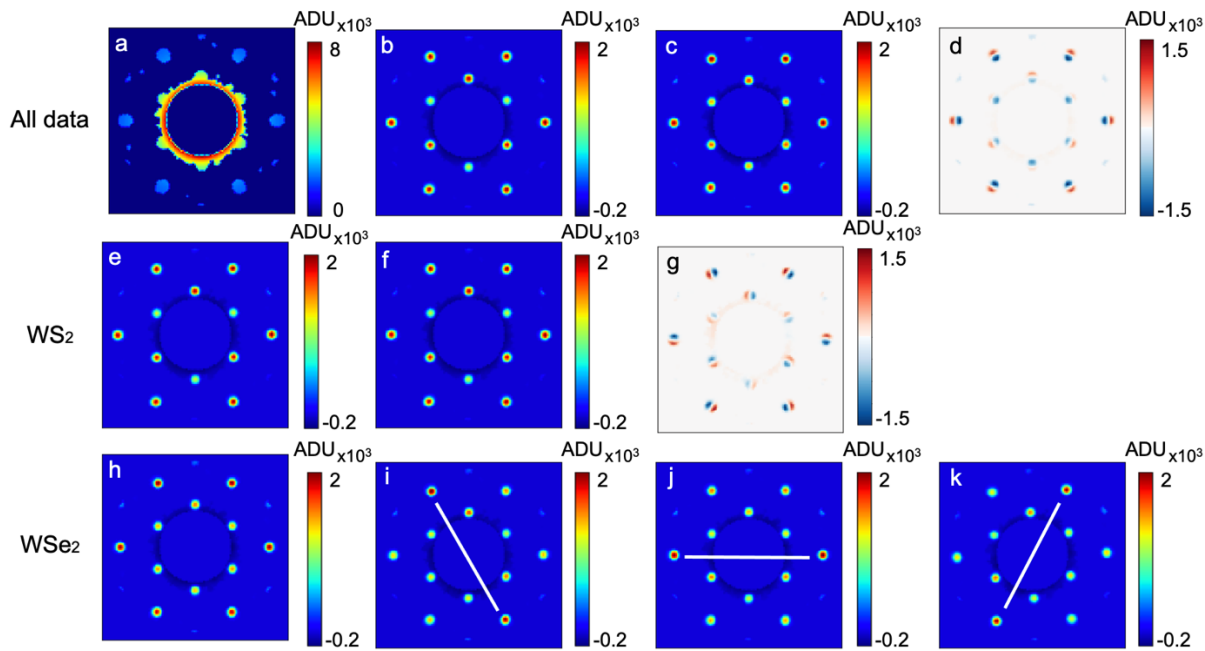

**Figure S5. Mean diffraction patterns of each cluster for the junction sample in Fig. 2.** (a) Mean diffraction patterns of the amorphous substrate. (b-c, e-f, h-k) are cluster centers subtracted by the background to show the diffraction spots with less noise. (b-c) Mean diffraction patterns of $WS_2$ and $WSe_2$; (d) Lattice constant difference between (b) and (c). (e-f) Mean diffraction patterns of sub-clusters in the $WS_2$ sample. (g) Rotational difference between (e) and (f). (h-k) Mean diffraction patterns of sub-clusters in the $WSe_2$ sample. (h) shows the cluster mean of the flat area in $WSe_2$, while (i), (j) and (k), present the cluster means of three directional ripple areas.

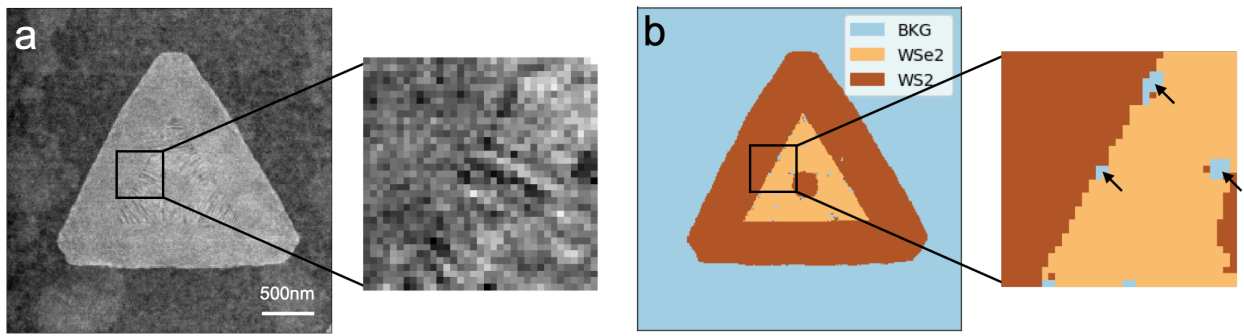

**Figure S6. Holes identified in the WS$_2$-WSe$_2$ junction sample by unsupervised learning.** (a) ADF image of the WS$_2$-WSe$_2$ junction with a zoomed-in area displayed. (b) Real-space map of the clustering results achieved using our method, where a few holes are identified in the WSe$_2$ (indicated with arrows). Comparing to the ADF image in (a), these holes are probably from the damage of local defects or dislocations in the material during the growth or annealing process.

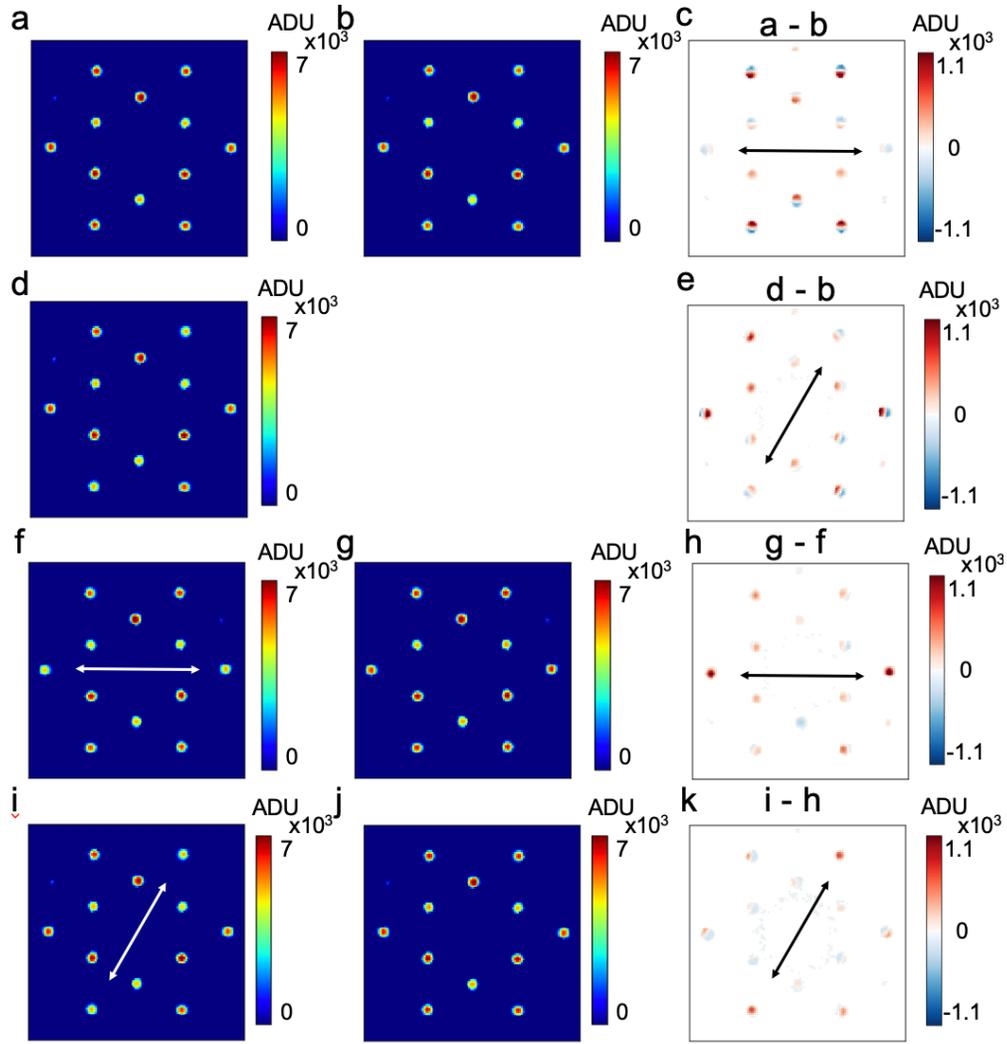

**Figure S7. Diffraction pattern difference between different clusters in WS$_2$-WSe$_2$ superlattices.** (a, d) Mean diffraction patterns of strain 1 and strain 2 cluster in Figure 3d. (c) Mean diffraction pattern of WS$_2$ region in Figure 3c. (c,e) Diffraction pattern difference between each sub-cluster of WSe$_2$ in Fig. 3d and the WS$_2$ area, indicating a lattice-matched coherent interface between the two materials along the direction pointed by arrows. (f, g, i, j) Mean diffraction patterns of Ripple 1 (f), Flat Region 1 (g), Ripple 2 (i) and Flat Region2 (j) in Figure 3e. (h,k) Diffraction patterns difference between the sub-clusters in Fig. 3e, where (h) represents Ripple 1 minus Flat 1, while (k) represents Ripple 2 minus Flat 2. The intensity difference of diffraction spots between sub-clusters indicates a ripple (or tilt) in the lattice, which is consistent with our observation in the junction sample in Fig. 2 and our previous work on the superlattice[2]. The tilt axes are labeled by the white lines.

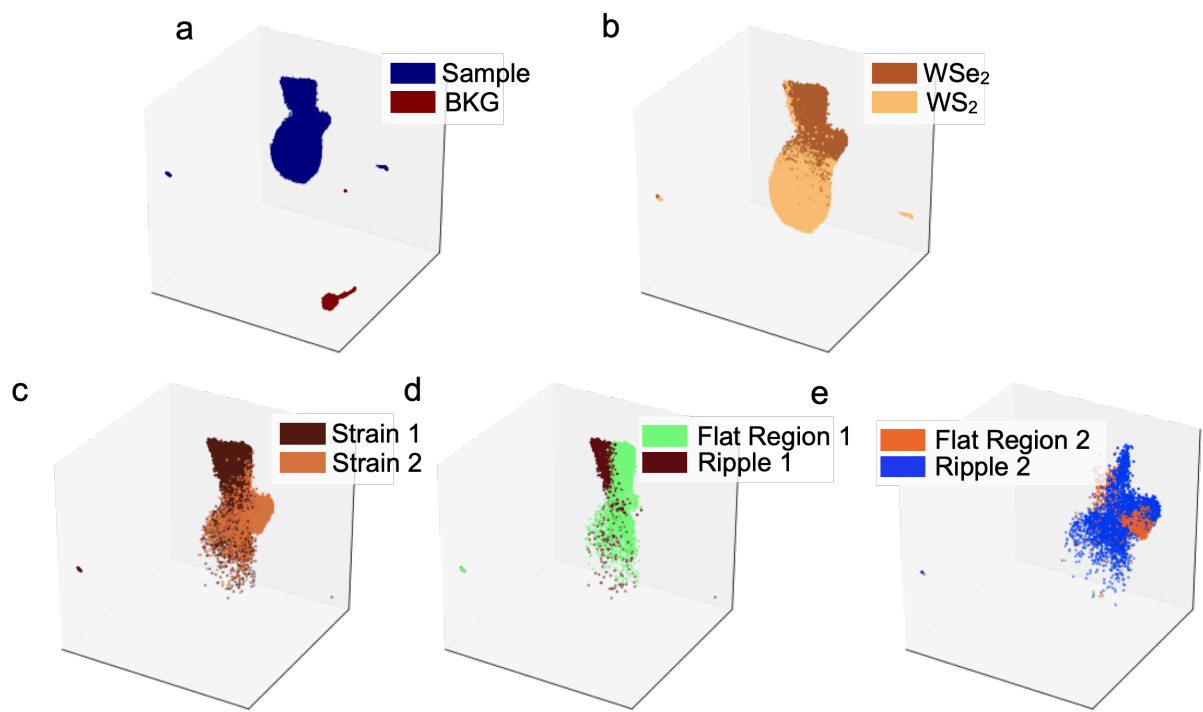

**Figure S8. Manifold structures of WS$_2$-WSe$_2$ superlattices.** (a) 3D manifold structure of the data in the first-round clustering. (b) Manifold structure of the sample subcluster (including WS$_2$ and WSe$_2$) in the second-round clustering. (c) Manifold structure of WSe$_2$ displaying uniaxial strain in the third round. (d,e) Manifold structures of the final-round data uncovering minor ripples in the sample.

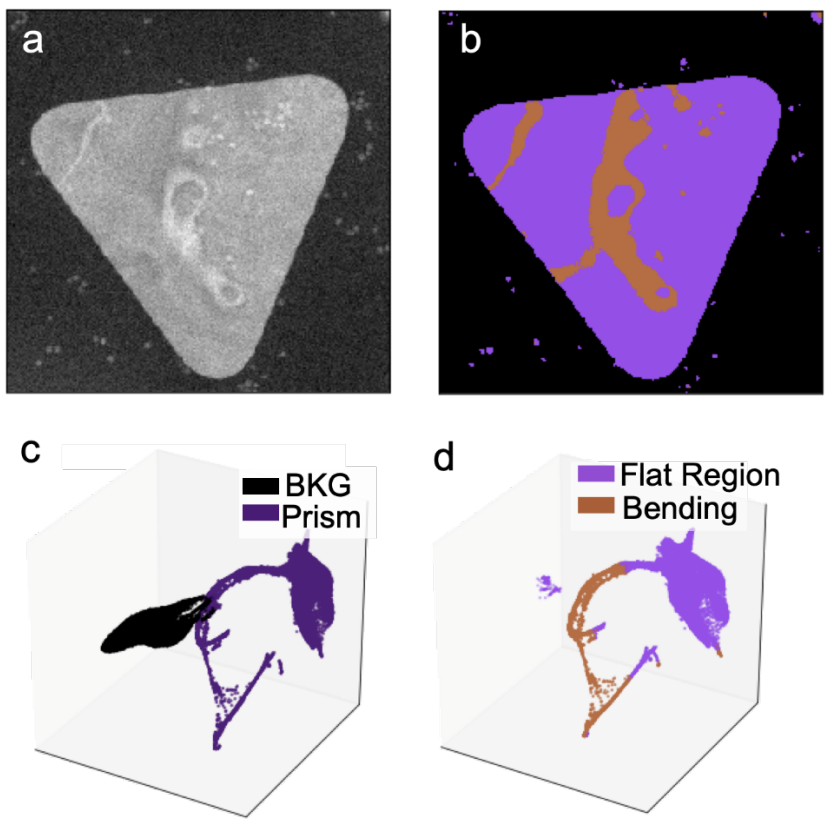

**Figure S9. Clustering results of an entire silver nanoprism.** (a) ADF-STEM image of the entire silver nanoprism flake. (b) Real-space map of the clustering results, showing bending contours in the nanoprism (purple). (c-d) 3D manifold structure of the data in different rounds.

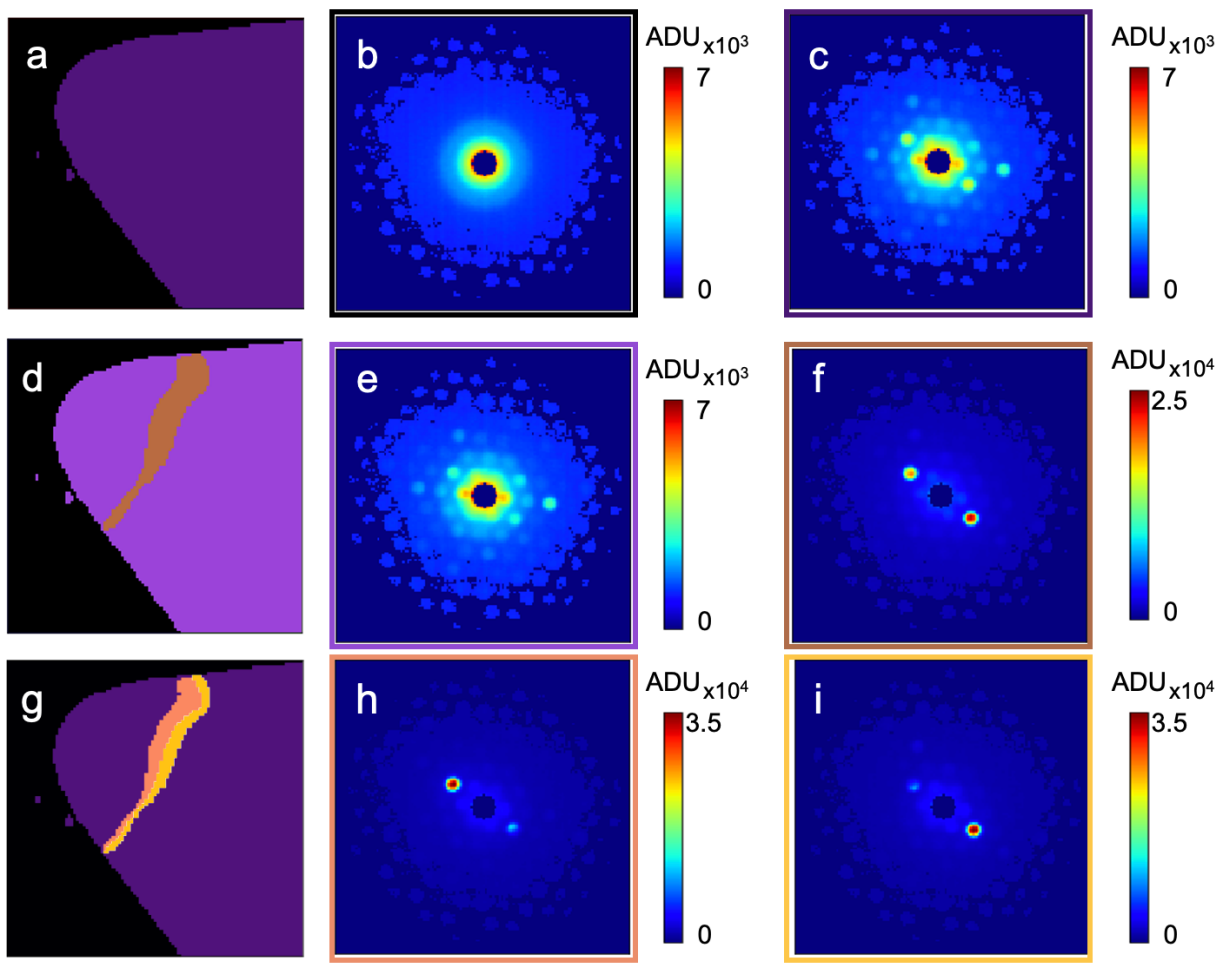

**Figure S10. Clustering results of silver nanoprism.** (a) Real-space map of the first-round clustering result, where the Ag nanoprism is distinguished from the background. (b-c) Mean diffraction pattern of background (b) and Ag nanoprism (c). (d) Real-space map of the second-round clustering result, where the bending contour (brown) and the flat area (purple) in the Ag nanoprism are separated. (e-f) Mean diffraction pattern of mean flat region (e) and bending contour region (f). (g) Real-space map of the third-round clustering result, where the bending contour are separated into two sides. (h-i) Mean diffraction pattern of two sides of the bending contour.

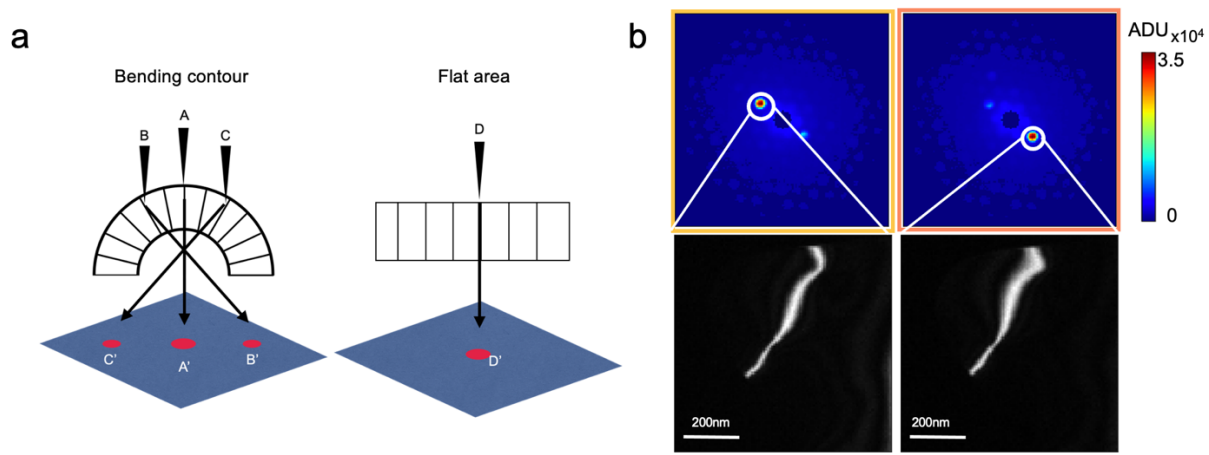

**Figure S11. Clustering results of silver nanoprisms.** (a) Schematic of the electron diffraction at a bending contour and a flat area. When the electron probe scans the two sides of the bending contour (B and C), the angles fulfill different Bragg conditions, leading to high intensity in different diffraction spots (B' and C' respectively). When the probe scans a flat area in the Ag nanoprism (A or D), most of the intensity lies in the unscattered bright disk (A' and D') as the atomic plane is parallel to the incident beam. (b) Virtual dark-field images of both sides of the bending contour from the 4D dataset by placing virtual apertures on the corresponding diffraction spots.

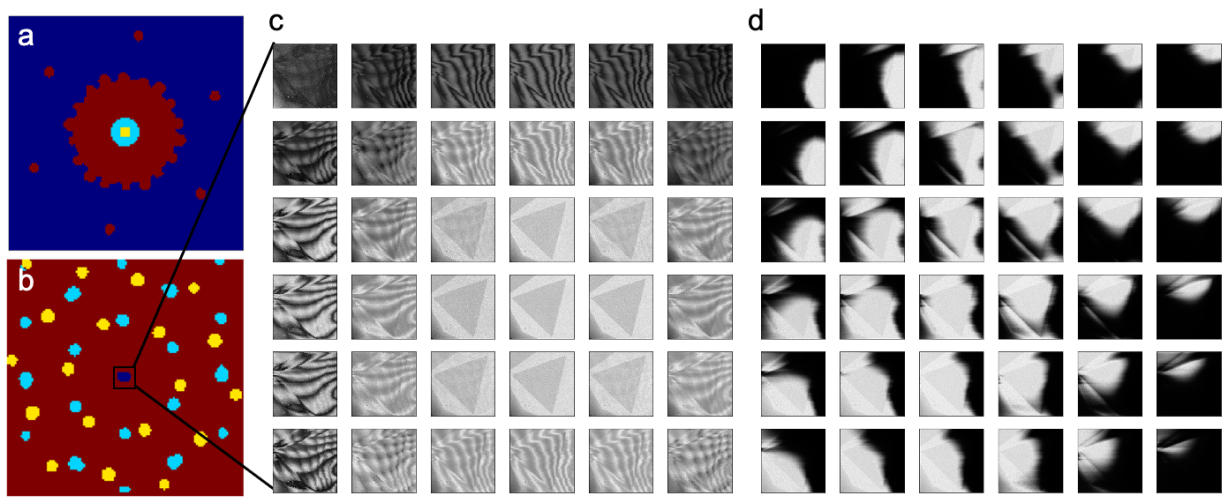

**Figure S12. Clustering real-space images in 4D datasets.** (a) Map of the clustering results in diffraction space without normalization image intensity in the pre-processing. The data are clustered according to the intensity rather than actual features in the sample. (b) Map of the clustering results in diffraction space with real-space normalization, which segment the diffraction space into different areas according to the virtual imaging mode (BF or DF) or the crystallinity in the sample. (c) Pixel-by-pixel real space images from the center beam, where we observed stripes in the BF images at the edge of the center diffraction disk. (d) Pixel-by-pixel real space images from the center beam of the raw data without any pre-processing, indicating the stripes we observed are artifacts caused by the diffraction pattern alignment.

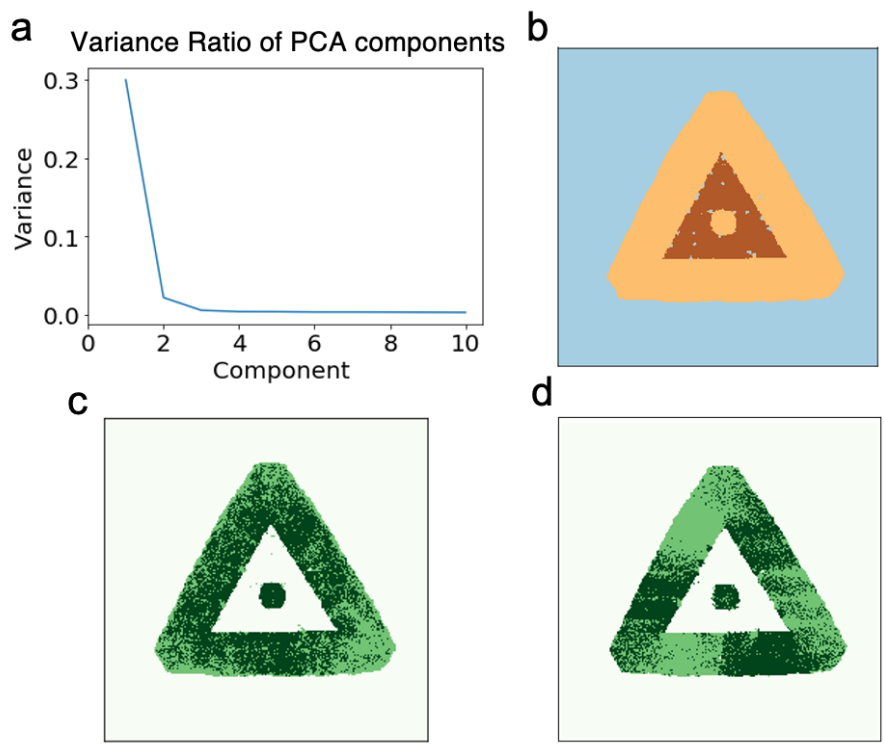

**Figure S13. PCA and UMAP Failure Cases** (a) Variance Ratio of PCA components from the whole $WS_2$-$WSe_2$ heterojunction dataset shown in Figure 2. Based on the curve we choose the first three high variance components to do the clustering. (b) First and second round clustering results based on PCA able to separate background, $WS_2$, and $WSe_2$. (c) Third round cluster labels from PCA. The components from PCA only capture the main feature but it cannot extract the minor features, such as the small rotation angles in the $WS_2$ sample. (d) Third round cluster labels from UMAP can partially capture the features, but introduces striping artifacts.